\documentclass[aps,prb,superscriptaddress,twocolumn,final,floatfix,showpacs]{revtex4-1}

\pdfoutput=1

\usepackage{amsmath}
\usepackage{graphicx}
\usepackage{latexsym}
\usepackage{units}

\begin{document}

\title{First-principles modeling of the thermoelectric properties of 
       SrTiO$_3$/SrRuO$_3$ superlattices}

\author{ Pablo Garc\'{\i}a-Fern\'andez }
\affiliation{ Departamento de Ciencias de la Tierra y
              F\'{\i}sica de la Materia Condensada, Universidad de Cantabria,
              Cantabria Campus Internacional,
              Avenida de los Castros s/n, 39005 Santander, Spain}

\author{ Marcos Verissimo-Alves }
\affiliation{ Departamento de Ciencias de la Tierra y
              F\'{\i}sica de la Materia Condensada, Universidad de Cantabria,
              Cantabria Campus Internacional,
              Avenida de los Castros s/n, 39005 Santander, Spain}

\author{ Daniel I. Bilc }
\affiliation{ Physique Th\'eorique des Mat\'eriaux, 
              Universit\'e de Li\`ege, All\'ee du 6 de Ao\^ut 17 (B5), 
              B-4000 Sart Tilman, Belgium}
\affiliation{ Molecular and Biomolecular Physics Department, 
              National Institute for Research and Development of Isotopic
              and Molecular Technologies, RO-400293, Cluj-Napoca, Romania}

\author{ Philippe Ghosez }
\affiliation{ Physique Th\'eorique des Mat\'eriaux, 
              Universit\'e de Li\`ege, All\'ee du 6 de Ao\^ut 17 (B5), 
              B-4000 Sart Tilman, Belgium}

\author{ Javier Junquera }
\affiliation{ Departamento de Ciencias de la Tierra y
              F\'{\i}sica de la Materia Condensada, Universidad de Cantabria,
              Cantabria Campus Internacional,
              Avenida de los Castros s/n, 39005 Santander, Spain}

\date{\today}

\begin{abstract}
 Using a combination of first-principles simulations, 
 based on the density functional theory, and Boltzmann's semiclassical theory,
 we have calculated the transport and thermoelectric 
 properties of the half-metallic two dimensional
 electron gas confined in single SrRuO$_3$ layers of 
 SrTiO$_3$/SrRuO$_3$ periodic superlattices. 
 Close to the Fermi energy we find that the semiconducting 
 majority spin channel displays a very large in-plane component of the
 Seebeck tensor at room temperature,
 $S$ = 1500 $\mu$V/K, 
 and the minority spin channel shows good in-plane conductivity 
 $\sigma$ = 2.5 (m$\Omega$cm)$^{-1}$. 
 However, contrary to the 
 expectation of Hicks and Dresselhaus model about
 enhanced global thermoelectric properties due to the 
 confinement of the metallic electrons, we find that the total 
 power factor and thermoelectric figure of merit
 for reduced doping is too small for practical applications. 
 The reason for this failure can be traced back on the electronic
 structure of the interfacial gas, which departs from the 
 free electron behaviour on which the model was based.
 The evolution of the electronic structure, electrical conductivity,
 Seebeck coefficient, and power factor as a function of the
 chemical potential is explained by a 
 simplified tight-binding model. 
 We find that the electron gas in our system is composed by 
 a pair of one dimensional electron gases orthogonal to each other. 
 This reflects the fact the physical dimensionality of the electronic system 
 can be even smaller than that of the spacial confinement of the carriers.
\end{abstract}


% 72.20.Pa   Thermoelectric and thermomagnetic effects
% 73.40.-c   Electronic transport in interface structures
% 74.70.Pq   Ruthenates
% 73.20.At   Surface states, band structure, electron density of states

\maketitle

%%% Introduction
\section{Introduction}
% Importance of thermoelectrics

 The interest in thermoelectric materials has undergone a revival over the 
 last decade.~\cite{Tritt-06,Zheng-08} 
 The reason behind this resurgence is double. 
 On the one hand, recent 
 experimental~\cite{Ohta-07.2,Dragoman-07,Venkatasubramanian-01,Kim-06} 
 and theoretical~\cite{Hicks-93,Hicks-93.2,Mahan-96,Slack-95} discoveries 
 allowed to glimpse new ways to significantly improve the efficiency 
 of these materials, quantified by the thermoelectric 
 adimensional figure of merit 

 \begin{equation}
    ZT=\frac{S\sigma^2}{\kappa_{\rm e} + \kappa_{\rm l}}T,
    \label{eq:ZT}
 \end{equation}

 \noindent where $S$ is the Seebeck coefficient (also called thermopower and
 denoted by $\alpha$ by some authors), 
 $\sigma$ is the electrical conductivity, 
 $\kappa_{\rm e}$ ($\kappa_{\rm l}$) is the electronic (lattice) contribution
 to the thermal conductivity, and $T$ is the absolute temperature. 
 On the other hand, there is an indubitable technological 
 interest over these systems, as significant improvements in this field will 
 potentially impact fuel consumption efficiency and allow to build 
 diminute cooling devices without moving parts.~\cite{Tritt-06} 

 It is usually accepted that applications in this field will only be 
 cost effective for materials where $ZT$ is significantly larger than 1. 
 However, the task of increasing its value remains challenging,
 since all the participating material's parameters in Eq.~(\ref{eq:ZT}) 
 are strongly interconnected, and also dependent on 
 material's crystal structure, electronic structure and carrier
 concentration.~\cite{Madsen-06}
 Increasing the Seebeck coefficient for simple materials
 requires a decrease in the carrier concentration, that yields to 
 a concomitant reduction in the electrical conductivity. 
 Also, an increase in the electrical conductivity leads to a comparable 
 increase in the electronic contribution to the thermal conductivity
 (as expressed in the Wiedemann-Franz law.) 
 In conventional solids, a limit is rapidly obtained 
 where a modification in any of these parameters adversely affects 
 other transport coefficients so that the resulting $ZT$ for a given material 
 at a given temperature does not vary significantly.~\cite{Hicks-96} 
 
 Typical good thermoelectrics are doped semiconductors with intermediate
 values of the carrier density (close to 10$^{18}$-10$^{19}$ cm$^{-3}$)
 displaying, at the same time, large Seebeck coefficients 
 and good charge mobility. 
 A typical example~\cite{Tritt-06} of this behavior is Bi$_2$Te$_3$ that,
 after doping, displays a $ZT$ value close to 1 at 320~K
 with a resitivity $\approx 1$ m$\Omega$cm and a Seebeck coefficient of 
 225 $\mu$V/K. 
 Other heavy-metal-based materials, such as group IV chalcogenides  
 (lead telluride, PbTe, and related materials), 
 exhibit large figure of merit at intermediate temperatures (up to 850 K).
 More interesting materials such as 
 clathrates,~\cite{Saramat-06} half-Heusler,~\cite{Shen-01,Culp-08} 
 skutterudites,~\cite{Sales-96,Singh-97,Keppens-98} 
 or strongly correlated oxides.~\cite{Matsubara-01,Ohta-05,
 Ohta-08,Wang-09.2,Hebert-10} 
 have also been identified.
 Finally, it is noteworthy that graphene layers,~\cite{Dragoman-07}
 silicon nanowires~\cite{Hochbaum-08,Vo-08} and molecular junctions 
 have also received much attention due to their low thermal conductivity.

 A completely new route to enhance the figure of merit was opened by 
 Hicks and Dresselhaus in the early nineties.~\cite{Hicks-93} 
 With a theoretical model, based on a simplified electronic band structure
 (assuming free-electron parabolic bands in two dimensions and a localized
 bound state in the third direction), these authors showed that a significant 
 increase in $ZT$ would be possible due to the modification 
 of the electronic properties of some materials when prepared in 
 the form of quantum-well superlattices ~\cite{Hicks-93} 
 or nanowires.~\cite{Hicks-93.2}
 In particular they predicted that production of bidimensional 
 heterostructures has two main consequences that improve the value of $ZT$: 
 (i) The confinement of the charge carriers in a plane  
 is expected to reduce the dispersion of the 
 density of states (DOS) of the bulk system, creating a sharper DOS at 
 Fermi energy, and
 (ii) the nanostructuration of the system along a given spatial
 direction also favors the  
 dispersion of phonons that have wavelengths longer than the 
 period of the superlattice, leading to a decrease on the thermal 
 conductivity $\kappa_{\rm l}$, 
 due to scattering of the lattice vibrations by the interface 
 between layers.
 The first condition is in line with Mahan and Sofo's 
 proposal~\cite{Mahan-96} for an ``ideal thermoelectric".
 From a purely mathematical point of view, these authors 
 found how a $\delta$-shaped transport distribution 
 maximizes the thermoelectric properties.

% Experimental results confirming Hicks and introduction of the 2DEG

 Experimentally, these ideas have been thoroughly checked in 
 semiconducting heterostructures containing heavy non-metal ions.
 As a proof of concept, pioneering experimental works were carried
 out by Hicks and coworkers~\cite{Hicks-96} in PbTe/Pb$_{1-x}$Eu$_{x}$Te
 multiple quantum wells, showing a good agreement between the 
 experimental results and the theoretical model predictions for
 the increase in the figure of merit.
 Values of $ZT$ as high as 2.4 have been measured in $p$-doped 
 Bi$_2$Te$_3$/Sb$_2$Te$_3$ superlattices.~\cite{Venkatasubramanian-01} 
 However, these systems are far from ideal as they decompose at the 
 temperature where thermoelectric materials are expected to 
 function (they start to decompose at $T \approx$ 200° C), 
 and contain poisonous elements like lead or bismuth.

 Also, in good agreement with the 
 Hicks and Dresselhaus prediction,
 Ohta {\it et al.}~\cite{Ohta-07.2} found that periodic SrTiO$_{3}$/Nb-doped 
 SrTiO$_{3}$ superlattices,
 where a two-dimensional electron gas (2DEG) is formed at the  
 Nb-doped layer, exhibit enhanced Seebeck coefficients as the width of the 
 doping layer is reduced. In particular, when its thickness 
 reaches the ultimate thickness of one unit cell
 a very high value of 
 $S$ = 850 $\mu$V/K was observed, and a $ZT$
 of 2.4 for the 2DEG was estimated (this corresponds to an effectice $ZT$
 of 0.24 for the complete device having the 2DEG as the active part).
 However, this enhancement of the thermoelectric properties in 
 2DEG at oxide superlattices seems not to be so universal.
 Recent experimental results~\cite{Pallecchi-10} could not find 
 any enhacement of the
 Seebeck effect due to the electronic confinement in the metallic 
 state at the $n$-type LaAlO$_{3}$/SrTiO$_{3}$ interface.~\cite{Ohtomo-04}

% STO-SRO
 
\begin{figure} [t]
    \begin{center}
        \includegraphics[width=\columnwidth]{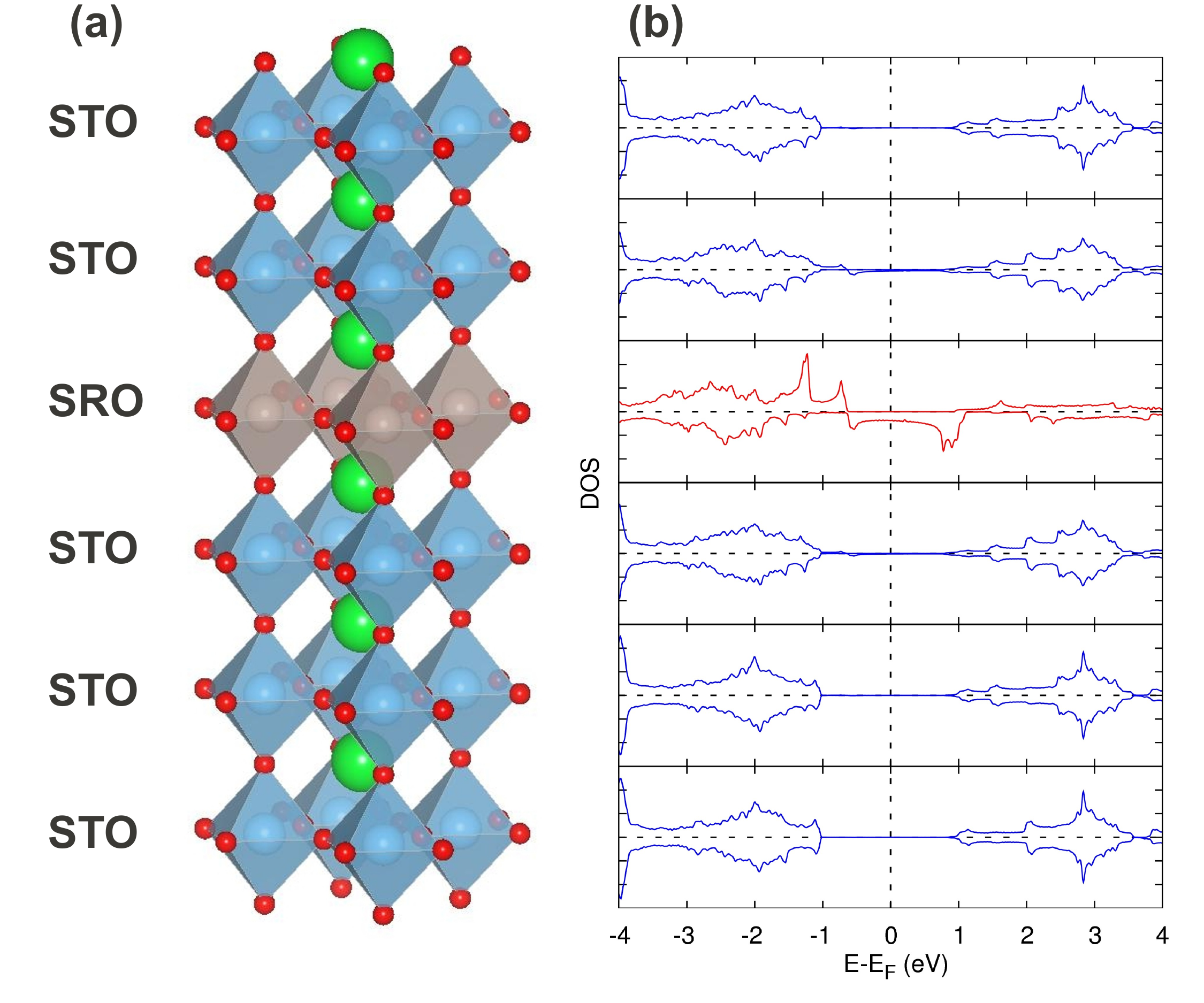}
        \caption{ (Color online) 
                  (a) Schematic representation of the unit cell, 
                  periodically repeated in space, of
                  (SrTiO$_{3}$)$_5$/(SrRuO$_{3}$)$_1$ superlattices.
                  Ti atoms (in blue)
                  Ru ones (in gray) are situated at the center of the 
                  octahedra, as denoted
                  by the layer labeling, while
                  O (in red) is placed at the vertex and 
                  Sr (in green) at the interstices.
                  (b) Layer-by-layer PDOS on the
                  atoms at the Sr$B$O$_{3}$ ($B$ = Ti or Ru) for the 
                  corresponding layer at the same height as in (a).
                  Majority (minority) spin is represented in the upper (lower)
                  half of each panel.
                  }
        \label{fig:super-dos}
    \end{center}
 \end{figure}

 Very recently, we have proposed an alternative mechanism to generate 2DEG  
 at oxide interfaces,~\cite{Verissimo-12} playing
 with the possibility of generating a quantum well based on the different
 electronegativity of the cations in the perovskite structures used
 to build the interface.
 In particular a half-metallic spin-polarized 2DEG 
 was theoretically predicted in 
 (SrTiO$_{3}$)$_{5}$/(SrRuO$_{3}$)$_{1}$ superlattices.
 The electron gas is fully localized in the SrRuO$_{3}$ layer due to
 higher electronegativity of Ru$^{4+}$ ions compared to Ti$^{4+}$ ones.
 The 2DEG presents a magnetic moment of $\mu=2 \mu_{\rm B}$/Ru ion. 
 All these properties can be seen in the layer-by-layer 
 projected density of states (PDOS)
 shown in Fig.~\ref{fig:super-dos}, 
 where the only contribution to the DOS at the Fermi energy comes from the 
 minority spin SrRuO$_{3}$ layer, while the majority spin 
 channel behaves like a 
 wide-gap semiconductor. 
 From the point of view of the 
 design of thermoelectric materials [see Eq.~(\ref{eq:ZT})] this 
 system seems very promising due to the possible combination of a 
 high-Seebeck coefficient (coming from the semiconducting majority
 spin channel) and metallic conductivity (coming from the 
 minority spin channel).

% Objetives

 The main objective of this work is to assess the thermoelectric properties 
 of the 2DEG in (SrTiO$_{3}$)$_5$/(SrRuO$_{3}$)$_1$ superlattices,
 and to ascertain whether the Hicks and Dresselhaus model is applicable
 or not in this kind of systems. 
 In order to achieve this goal, and starting from accurate
 first-principles electronic structure simulations on this interface,
 we use the Boltzmann transport 
 theory~\cite{Ziman, Ashcroft} within the constant scattering time
 approximation to obtain the Seebeck coefficient and other transport functions.

 The rest of the paper is organized as follows.
 Computational details are summarized in Sec.~\ref{sec:comp}.
 The main results on the Seebeck coefficient, electrical conductivity,
 and power factor for the superlattice are presented in Sec.~\ref{sec:res}.
 Finally, in Sec.~\ref{sec:dis} we discuss a simple model to understand the 
 physical origin of the previous transport coefficients.

%%% Computational details

\section{Computational details}
\label{sec:comp}

 We estimated the electrical conductivity and the Seebeck coefficient
 through the semiclassical Boltzmann theory within the constant 
 relaxation time approximation, as implemented in the {\sc BoltzTraP} 
 code.~\cite{Boltztrap}
 This implementation relies on the Fourier expansion of the band-energies,
 provided by a first-principles electronic structure code.
 Following our previous work, we have used {\sc siesta}~\cite{Soler-02} 
 to compute both the relaxed atomic and electronic band structures of 
 (SrTiO$_{3}$)$_5$/(SrRuO$_{3}$)$_1$ superlattice in the 
 local density approximation (LDA) to
 the density functional theory.
 An extra Hubbard-$U$ term,
 following the rotationally invariant LDA+$U$ scheme of
 Anisimov {\it et al.}~\cite{Anisimov-97},
 is included to account for the strong electron correlations,
 with a $U_{\rm eff}$ of 4.0 eV applied only to the $d$ orbitals of Ru, 
 as in Ref.~\onlinecite{Verissimo-12}.
 In order to get smooth Fourier expansion of the one-electron eigenenergies
 and converged transport coefficients we proceed in a two step procedure:
 (i) first we relax the atomic structure and the 
 one-particle density matrix with a 
 sensible number of $k$-points ($12 \times 12 \times 2$ 
 Monkhorst-Pack mesh~\cite{Monkhorst-76}), and 
 (ii) freezing-in the relaxed structure and density matrix, 
 we perform a non-self-consistent band 
 structure calculation with a much denser sampling of $72 \times 72 \times 17$  
 (5994 $k$-points in the irreducible 
 Brilloin zone; 88128 $k$-points in the full Brillouin zone).
 The rest of the computational parameters remain the same as in 
 Ref.~\onlinecite{Verissimo-12}.
 The robustness of the results presented below have been doubled-checked
 using the {\sc crystal09} code~\cite{Dovesi-05} within the 
 B1-WC hybrid functional~\cite{Bilc-08} that mixes the generalized 
 gradient approximation of Wu and Cohen~\cite{Wu-06.1} 
 with 16 \% of exact exchange with the B1 scheme.~\cite{Becke-96}

% Geometry 

 The atomic structure of the (SrTiO$_{3}$)$_5$/(SrRuO$_{3}$)$_1$ 
 superlattice at low temperature 
 includes the rotation of both the TiO$_6$ and RuO$_6$ octahedra along the 
 tetragonal $z$-axis of the system
 [to establish the notation, we will call the plane parallel to the interface
 the $(x,y)$ plane, whereas the perpendicular direction will be referred to
 as the $z$-axis].
 Taking into account that the temperature at which bulk SrTiO$_{3}$ undergoes
 the tetragonal to cubic transition is only 105~K, we expect that these 
 distortions are fully suppressed when the system is acting as a 
 thermoelectric generator at room or higher temperatures. 
 Therefore, in the present study we do not allow rotation and tiltings
 of the oxygen octahedra during the atomic relaxations of the superlattices.

% Transport calculations

 With the first-principles band structures,
 $\varepsilon_{i,{\bf k}}$, computed as indicated above,
 together with the space group symmetry of the superlattice, 
 we feed the {\sc BoltzTraP} code.
 There, after performing the Fourier expansion, 
 the conductivity tensor can be obtained as

 \begin{equation}
    \sigma_{\alpha \beta} (i, {\bf k}) = e^{2} \tau_{i,{\bf k}} 
                                         v_{\alpha} (i, {\bf k})
                                         v_{\beta}  (i, {\bf k}),
    \label{eq:condtensor}
 \end{equation}

 \noindent where $e$ is the electronic charge, $\tau_{i,\bf{k}}$ is 
 the relaxation time, and 

 \begin{equation}
    v_{\alpha} (i, {\bf k}) = \frac{1}{\hbar} 
                              \frac{\partial \varepsilon_{i,{\bf k}}}
                                   {\partial k_{\alpha}}
    \label{eq:groupvel}
 \end{equation}

 \noindent is the $\alpha$ component of the group velocity for an electron
 in band $i$.
 Now, from the previous conductivity matrix we can compute 
 the relevant transport tensors that relate the electric current with 
 an external electric field [$\sigma_{\alpha \beta}(T,\mu)$]
 or temperature gradients [$\nu_{\alpha \beta}(T,\mu)$].
 These tensors depend on the temperature, $T$, 
 and the chemical potential, $\mu$, 
 that determines the number of carriers
 or the level of doping.
 The final expressions are given by

 \begin{equation} 
    \sigma_{\alpha \beta}(T,\mu)= 
    \sum_i \int \frac{d {\bf k}}{8\pi^3} 
    \left[-\frac{\partial f (T,\mu)}{\partial \varepsilon}\right] 
    \sigma_{\alpha \beta} (i, {\bf k}),
    \label{eq:cond}
 \end{equation} 

 \noindent and

 \begin{equation}
    \nu_{\alpha\beta}(T,\mu) = 
    \frac{1}{T} \sum_i \int \frac{d {\bf k}}{8\pi^3}  
    \left[-\frac{\partial f (T,\mu)}{\partial \varepsilon}\right] 
    \sigma_{\alpha \beta} (i, {\bf k})
    [\varepsilon({\bf k})-\mu],
   \label{eq:trans} 
 \end{equation}

 \noindent where $f$ is the Fermi-Dirac distribution.
 Finally the components of the Seebeck tensor can be computed as 

 \begin{equation}
    S_{i j}(T,\mu)= \sum_{\alpha} \left(\sigma^{-1}\right)_{\alpha i}
                            \nu_{\alpha j}.
    \label{eq:see}
 \end{equation}

 The electronic contribution to the figure of merit is summarized in the 
 power factor, ${\rm PF} = S^{2}\sigma$, which is the numerator of 
 the right-hand side in 
 Eq.~(\ref{eq:ZT}). For a magnetic system the value of the PF 
 can be calculated from the individual spin bands 
 using Eq.~(\ref{eq:cond}) and Eq.~(\ref{eq:trans}) and

\begin{equation}\label{eq:PF}
   {\rm PF} = \left (\sigma_{\alpha\beta}^{\uparrow} + 
                     \sigma_{\alpha\beta}^{\downarrow} \right)^{-1} 
              \left (\nu_{\alpha\beta}^{\uparrow} + 
                     \nu_{\alpha\beta}^{\downarrow}\right)^2
\end{equation}

 While computing the previous transport properties two major approximations
 are considered: (i) the relaxation time $\tau$ is treated as constant,
 independent of temperature, band number, occupation and the $\bf k$ vector
 direction. While this approximation is fairly strong, 
 tests of this method\cite{Boltztrap} with semiconductor thermoelectrics like 
 Bi$_2$Te$_3$ and systems including electron correlation like 
 CoSb$_3$ lead to a reasonable agreement with experiment. 
 After Eq.~(\ref{eq:see}), this approximation allow us to compute the
 Seebeck coefficient on an absolute scale (independent of $\tau$).
 However the conductivity can be calculated only with respect the  
 relaxation time, and a value of $\tau$ has to be introduced as a parameter
 [typically using the theoretical $\sigma/\tau$ value obtained from 
 Eq.~(\ref{eq:cond})
 to reproduce exactly the experimental conductivity at a given temperature
 and carrier density $n$]. 
 Here, the relaxation time value 
 $\tau=0.43\times 10^{-14}$ s employed in the calculations was obtained from 
 fitting
 the room temperature conductivity $\sigma=1.667\times 10^5$ S/m of 
 bulk SrTiO$_3$ at
 electron concentration $n=1\times 10^{21}$ cm$^{-3}$,~\cite{Choi-10,Seo-07} 
 which is
 very similar to that obtained from SrRuO$_3$.~\cite{Chang-09,Cao-97}

 The second approximation is the ``rigid band approach''
 that assumes that the band structure does not change with temperature or
 doping, and therefore is fixed independently on the chemical
 potential.

%%% Results 

\section{Results}
\label{sec:res}

 \subsection{Electron localization and band structure}

% Confiment of the 2DEG and stability of the results

 In Fig.~\ref{fig:super-dos} we show the DOS of the 
 (SrTiO$_{3}$)$_5$/(SrRuO$_{3}$)$_1$ 
 superlattices projected layer-by-layer. 
 As previously discussed in 
 Ref.~\onlinecite{Verissimo-12}, the electronic structure displays a 
 half-metallic state where the conduction only takes place for the 
 minority spin electrons that are completely confined to the SrRuO$_{3}$ layer. 
 For the majority spin our LDA+$U$ calculations predict a gap of
 $\approx 1.4$ eV around the Fermi energy, which is typical for 
 wide-gap semiconductors.

\begin{figure} [h]
    \begin{center}
        \includegraphics[width=\columnwidth]{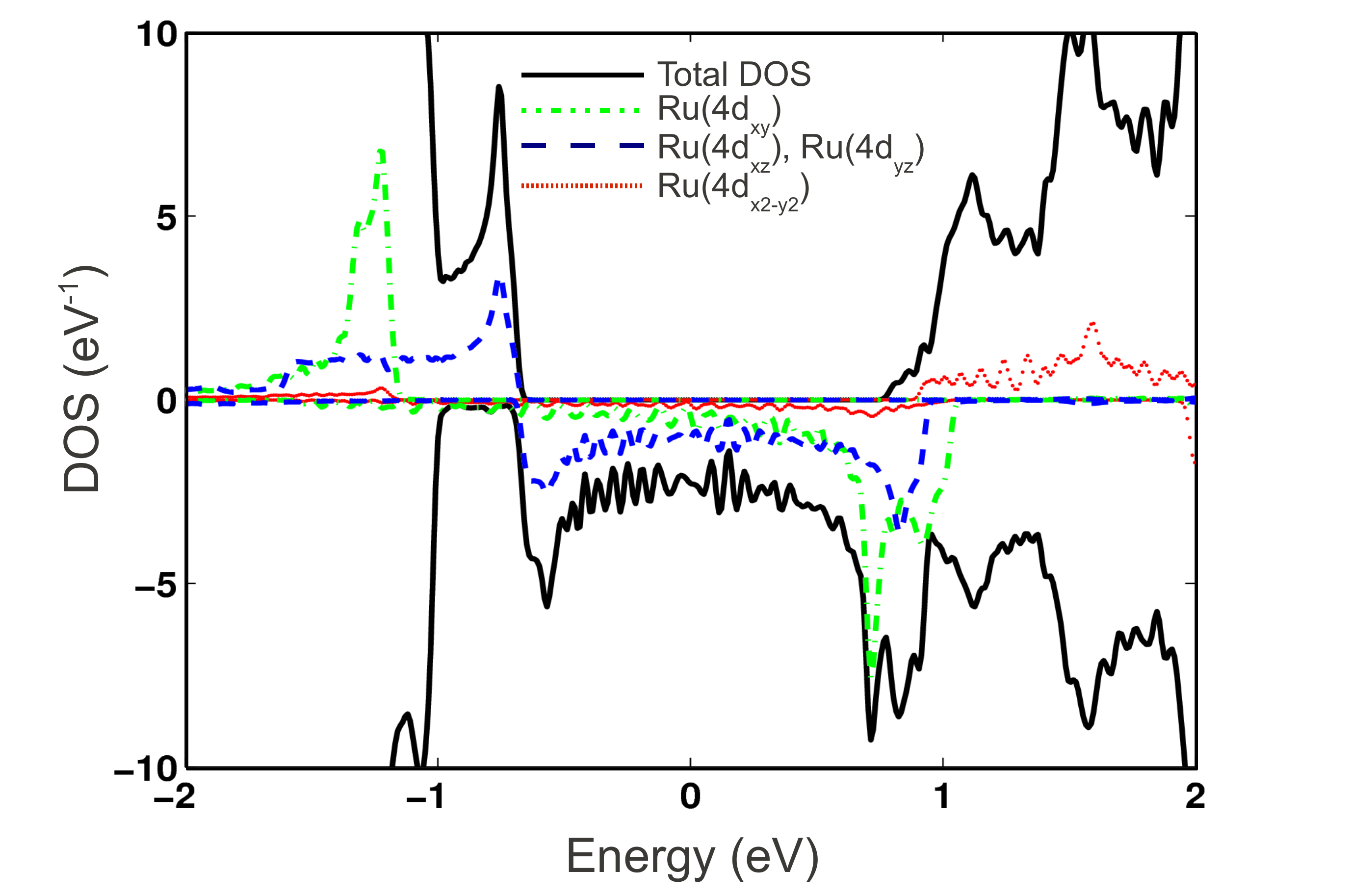}
        \caption{ (Color online) Total DOS (solid black lines) 
                  and PDOS on some Ru(4$d$) orbitals, 
                  showing the main character of the bands around the 
                  Fermi energy in
                  (SrTiO$_{3}$)$_5$/(SrRuO$_{3}$)$_1$ superlattices.
                  Green dot-dashed lines are the projections on the 
                  4$d_{xy}$ orbitals,
                  blue dashed lines on the 4$d_{xz,yz}$ orbitals
                  (these curves are exactly degenerate),
                  and red dotted lines on the 4$d_{x^{2}-y^{2}}$
                  orbitals. 
                  The zero of energies is aligned at the Fermi level.
                  The wiggles around zero are caused by the finite $k$ 
                  resolution.}
        \label{fig:tb}
   \end{center}
 \end{figure}

 A more detailed description of the nature of the states around the 
 Fermi energy is obtained when we plot the PDOS
 for the 4$d$ orbitals of Ru (Fig. \ref{fig:tb}). 
 Since Ru$^{4+}$ is a low-spin $d^4$ transition metal ion, we expect 
 the conduction band to have a strong $t_{2g}$($4d_{xy}$,$4d_{xz}$,$4d_{yz}$) 
 character. 
 Indeed, the conduction band in the minority spin is formed by the half-filled 
 degenerate Ru($4d_{xz,yz}$) orbitals while the Ru($4d_{xy}$) band is 
 mainly situated at an energy slightly above the Fermi level. 
 Similarly, the majority spin valence band is composed by Ru $t_{2g}$ orbitals,
 but since they are shifted to lower energies they are strongly 
 mixed with O(2$p$) bands. 
 In Ref.~\onlinecite{Verissimo-12} these features were explained 
 with the use of a simplified tight-binding model including three 
 main physical ingredients: 
 (i) the bidimensionality of the SrRuO$_{3}$ layer, 
 (ii) the difference of in-plane and out-of-plane bonding for the Ru-ions,
 and (iii) the electron-electron interactions as described by a Hubbard term. 
 In Fig.~\ref{fig:tb} we can also see the Ru($4d_{x^2-y^2}$) band,
 however as it lies at relatively high-energies its presence is 
 negligible when discussing the transport properties in this system.

\subsection{Transport calculations}

\begin{figure} [h]
    \begin{center}
       \includegraphics[width=\columnwidth]{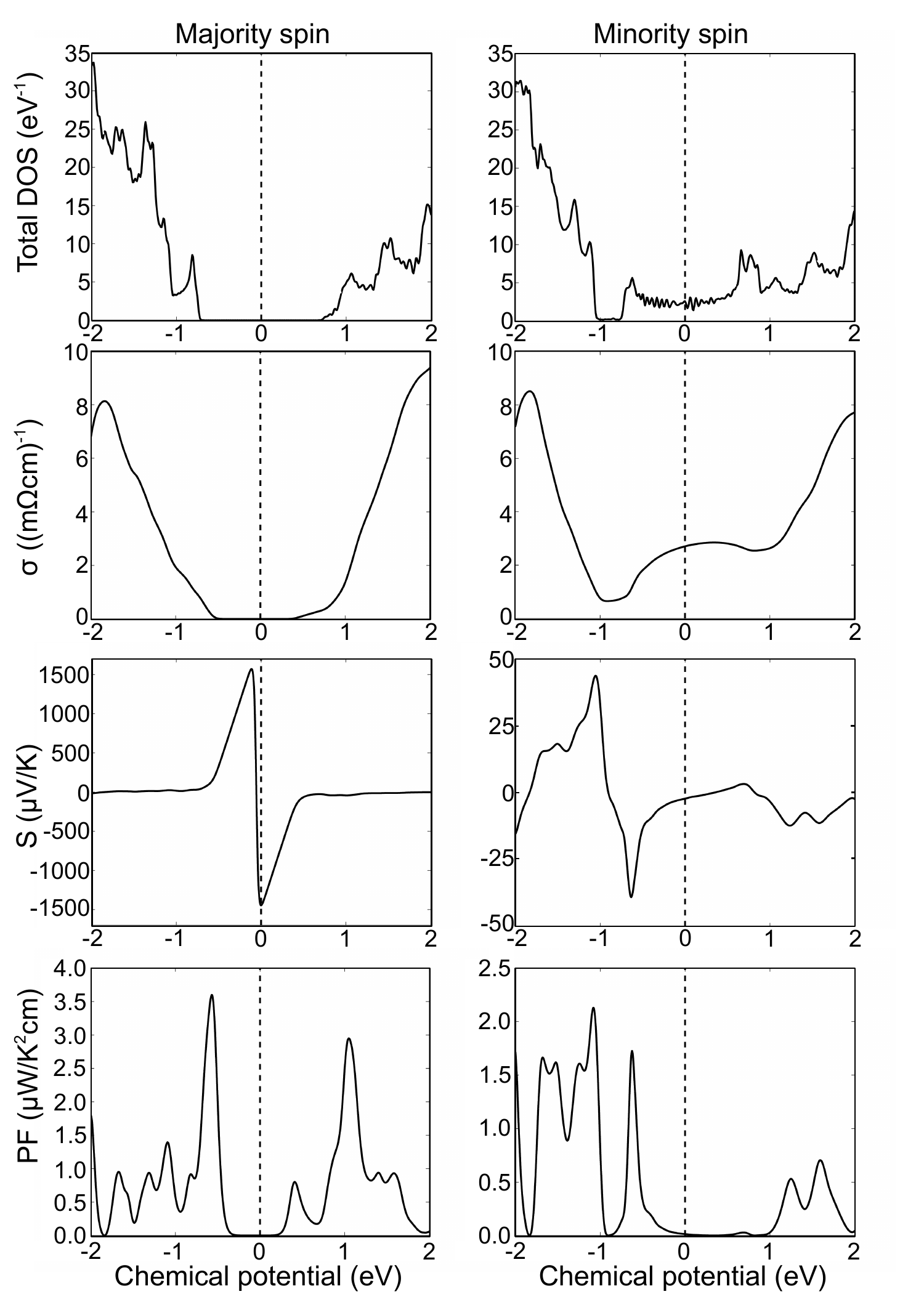}
       \caption{ Results of the calculation of the electronic
                 DOS and transport properties at 300 K 
                 [conductivity ($\sigma$), Seebeck coefficient ($S$),
                 and power factor (PF)]
                 of (SrTiO$_{3}$)$_5$/(SrRuO$_{3}$)$_1$ superlattices 
                 as obtained by 
                 Boltzmann's semiclassical transport theory. 
                 On the left column we show results for the majority spin bands
                 and on the right for the minority spin ones.}
        \label{fig:trans}
    \end{center}
 \end{figure}

 In Fig.~\ref{fig:trans} we show the calculated electrical conductivity,
 Seebeck coefficient and power factor for $T$ = 300 K as a function of 
 the position of the chemical potential (i.e. doping level) 
 for both the majority and minority spin channels.
 We also compare them to their corresponding DOS.
 Due to the tetragonality of the superlattice, all the previous transport
 tensors are diagonal with only two independent components: 
 one parallel to the interface ($xx$ = $yy$), and
 a second one perpendicular to the interface ($zz$ component).
 Since all the carriers are confined to move in the SrRuO$_{3}$ plane, 
 from now on we will only focus on the parallel one.

 As expected for a half-metal, the behavior of these quantities 
 around the Fermi energy at zero doping is very different according 
 to the different nature of each spin channel. 

% conductivity

 In the case of the minority spin, the system is metallic in the 
 SrRuO$_{3}$ layer (Fig. \ref{fig:tb}), and the conductivity presents
 a local maximum close to $\mu=0$. Then, $\sigma$ decreases quickly as 
 the chemical potential decreases and gets closer to the Ru (4$d_{xz,yz}$) 
 band edge
 (around -0.6 eV below the Fermi energy, see Fig.~\ref{fig:tb}).
 The other spin component (the majority one) is semiconducting and the small 
 non-zero contribution to $\sigma$ (indiscernible from zero in the
 scale of the figure) comes from the use of a finite temperature 
 in the simulation (electrons thermally excited to the conduction bands,
 leaving behind holes in the valence band). 
 In both cases the conductivity quickly increases when $\left|\mu\right|>1.5$ eV
 as the doping starts to involve the large density of states associated 
 with SrTiO$_{3}$ levels. 
 When we compare these values with those of typical high-efficiency 
 thermoelectrics, $\sigma\approx 1 (m\Omega cm)^{-1}$, we find that the value
 for the minority spin is larger [$\approx 3 (m\Omega cm)^{-1}$] while 
 that of the majority channel is much smaller.

% Seebeck

 Similarly, the curves for $S$ also reflects the different nature of 
 the spin components. 
 In the case of the majority spin we note a discontinuity of the 
 Seebeck coefficient, showing the change from hole 
 (region with $\mu < $ 0) to electron (region with $\mu >$ 0) doping. 
 The maximum absolute value for this channel is $S\sim$ 1500 $\mu$V/K and 
 occurs for relatively small hole dopings ($\mu=-0.25$ eV) 
 This value is significantly larger than 
 the room temperature $S$ = 480 $\mu$V/K found by Ohta 
 {\it et al.}~\cite{Ohta-07.2} for the 2DEG in one unit-cell-thick 
 Nb-doped SrTiO$_{3}$ superlattices, 
 than the value of $S$ = 1050 $\mu$V/K in TiO$_{2}$/SrTiO$_{3}$ 
 heterointerfaces with an electron concentration of 
 $7\times 10^{20}$ cm$^{-3}$,~\cite{Ohta-07.2}
 and also than those typical associated to good bulk 
 thermoelectrics, which are usually~\cite{Tritt-06} around 150-250 $\mu$V/K
 (the reader has to keep in mind that we are comparing \emph{maximum} values
 of the Seebeck coefficient, thought they can be achieved at different 
 carrier concentrations of holes or electrons.) 
 The opposite happens to the minority spin channel, where
 the Seebeck coefficient $S$ is very small at $\mu=0$ 
 as the conduction changes from being dominated by electrons to holes 
 in the semi filled Ru($4d_{xz,yz}$) bands. 
 When the system is doped, the absolute value increases linearly with the 
 chemical potential but the energy scale is two orders of magnitude smaller 
 than in the majority spin case, a typical factor when comparing the Seebeck 
 coefficients of metals and semiconductors. 
 Only when the chemical potential is close to the lower edge of the 
 conduction band ($\mu\sim-0.6$ eV) the minority spin shows a pronounced 
 enhancement of the thermopower reaching a moderate value of 
 $S$ = -35 $\mu$ V/K.

% PF

\begin{figure} [h]
    \begin{center}
       \includegraphics[width=\columnwidth]{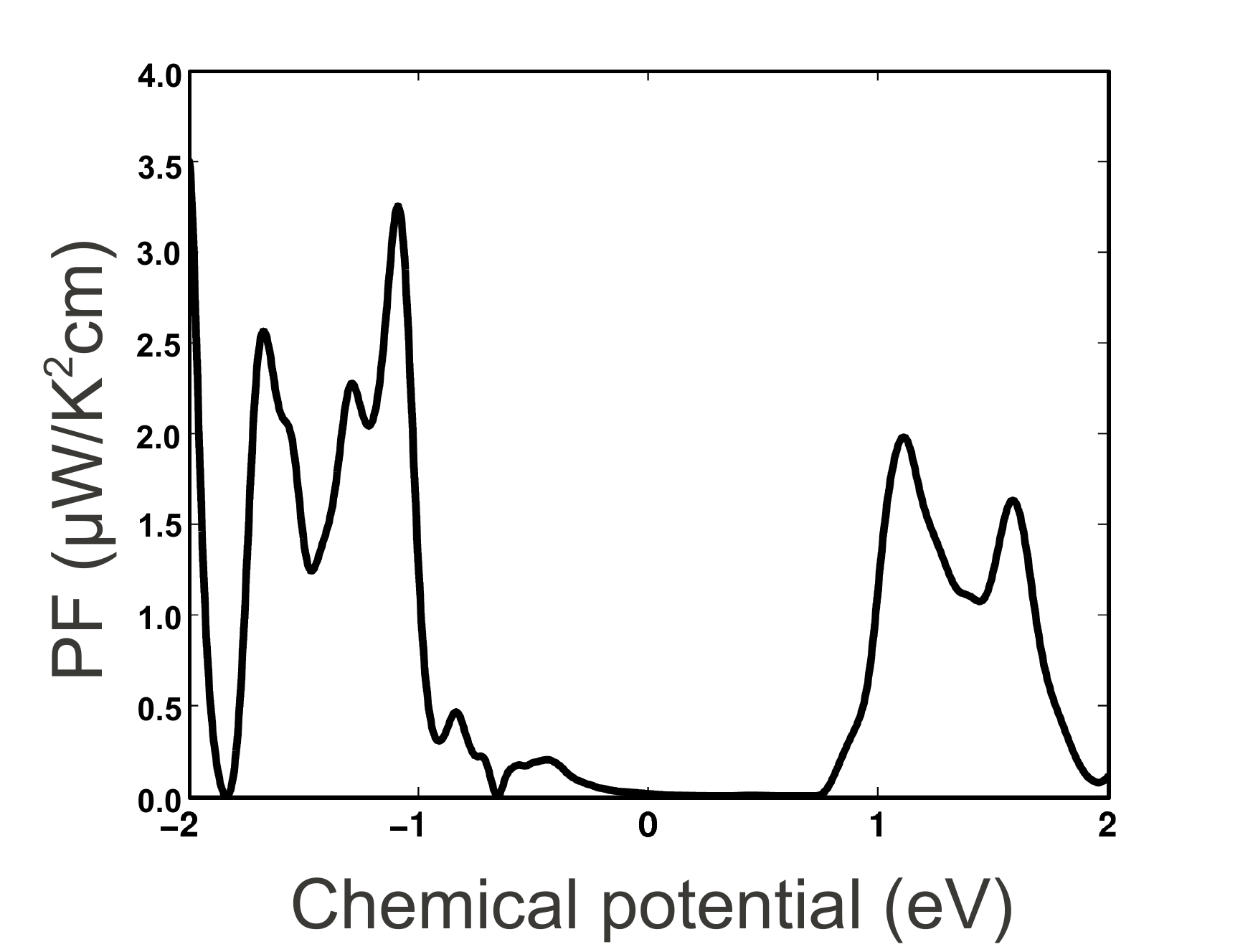}
       \caption{ Total PF for the (SrTiO$_{3}$)$_5$/(SrRuO$_{3}$)$_1$ 
                 superlattices at 300 K 
                 as a function of the chemical potential.}
        \label{fig:totalpf}
    \end{center}
 \end{figure}

 Calculation of the power factor for each of the spins shows strong compensation
 of Seebeck coefficient and conductivity in both channels 
 giving rise to very small 
 values ($<0.5$ $\mu$W/K$^2$cm) around the Fermi energy. 
 Indeed, while the majority spin displays a large Seebeck coefficient 
 and a very small conductivity, in the minority spin a reversed situation 
 is found.
 In both cases the power factor for the undoped system is almost 
 negligible when compared to those of good thermoelectrics 
 (20 - 50 $\mu$W/K$^2$cm). 
 Only for very strong hole doping, close to one hole per Ru$^{4+}$ ion, 
 an appreciable enhanment is observed for the power factor, both for the 
 majority spin component, PF=3.6 $\mu$W/K$^2$cm and the minority spin one, 
 PF=1.7  $\mu$W/K$^2$cm. 
 This result is summarized in Fig.~\ref{fig:totalpf}, where the total PF 
 is calculated as indicated in Eq.~(\ref{eq:PF}). 
 There we can observe that the total power factor is still 
 very small for low dopings, and only when the system is strongly 
 hole-doped ($>$1e/Ru) an appreciable PF is obtained.  

\section{Discussion}
\label{sec:dis}

 In the previous section we have found that even though 
 (SrTiO$_{3}$)$_5$/(SrRuO$_{3}$)$_1$ 
 superlattices exhibit a 2DEG, whose width is confined just within a 
 single perovskite unit cell, the power factor does not 
 display the strong enhancement expected by Hicks and 
 Dresselhaus.~\cite{Hicks-93}
 In order to understand the previous results we will analyze them using 
 both a free electron and a tight-binding model. 
 The choice of the free electron model was motivated by the fact that the
 Hicks and Dresselhaus works were based on parabolic bands to describe
 the electronic structure of the system, 
 while the tight-binding model has proven to be very successful to explain
 the first-principles results of the superlattice under study.

\begin{figure} [h]
    \begin{center}
       \includegraphics[width=\columnwidth]{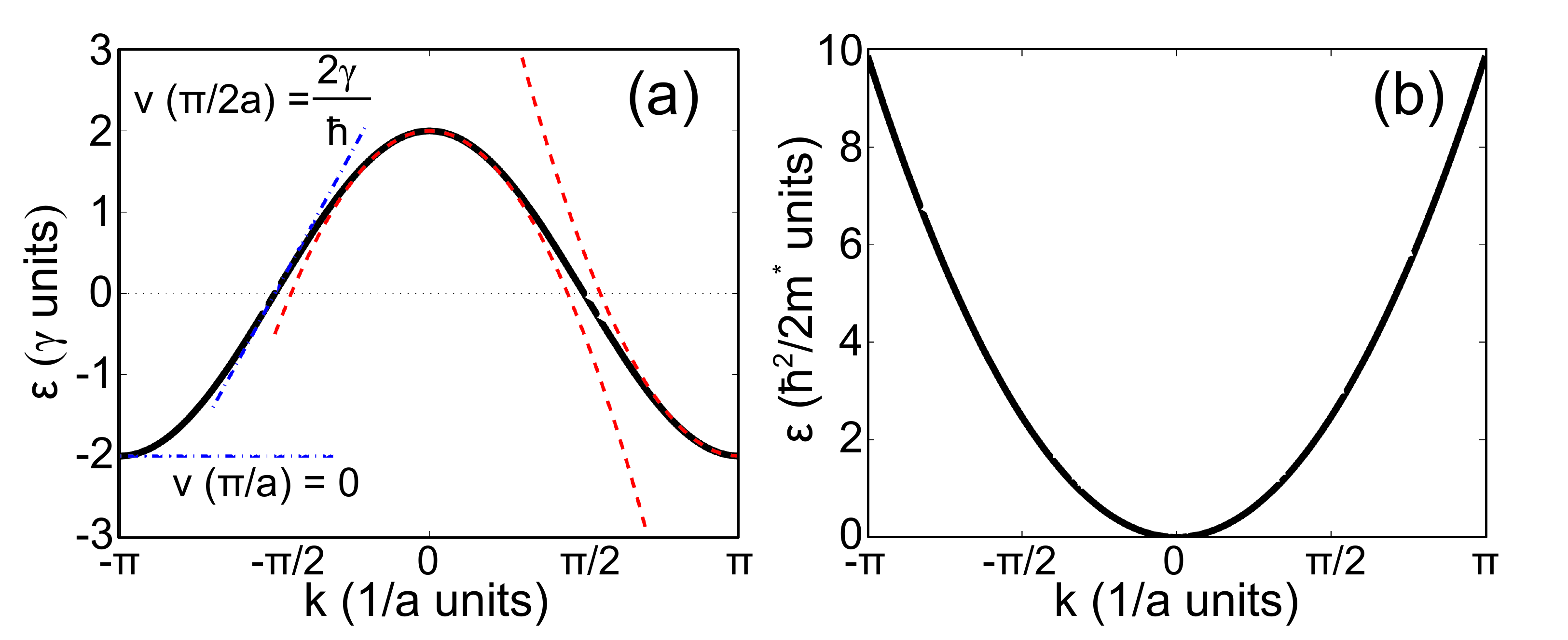}
       \caption{(Color online) Graphical representation of
                (a) one dimensional tight-binding, 
                and (b) free-electron bands that illustrate,
                respectively, Eq. (\ref{eq:tb}) and Eq. (\ref{eq:fe}). 
                In (a) we show with red dashed parabolae the free-electron 
                band approximation to the tight-binding model at the 
                band-edge while the blue dash-dotted line 
                depicts the group velocity
                at mid-band ($\varepsilon=0$) and the band-edge 
                ($\varepsilon=\pm2\gamma$).}
        \label{fig:bands}
    \end{center}
 \end{figure}

 We define a simple one band free electron model in one, two or 
 three dimensions using the energy dispersion,

 \begin{equation}
    \varepsilon({\bf k})=\frac{\hbar^2}{2m^\star} \sum_i k_i^2,
    \label{eq:fe}
 \end{equation}

 \noindent where $m^\star$ is the (isotropic) effective mass and  
 $i$ runs over the dimensions of the system. 
 Similarly, we define the energy dispersion for a one-band tight-binding
 model as

 \begin{equation}
    \varepsilon({\bf k})=2\gamma \sum_i \cos(k_i a),
    \label{eq:tb}
 \end{equation}

 \noindent where $\gamma$ is the characteristic interaction energy (band width)
 of one of the tight-binding center with its first-neighbors along 
 $\langle 100 \rangle$, and $a$ is the cubic lattice spacing. 
 A graphical representation of the free electron and tight-binding 
 bands is presented in Fig.~\ref{fig:bands}.

\begin{figure} [h]
    \begin{center}
       \includegraphics[width=\columnwidth]{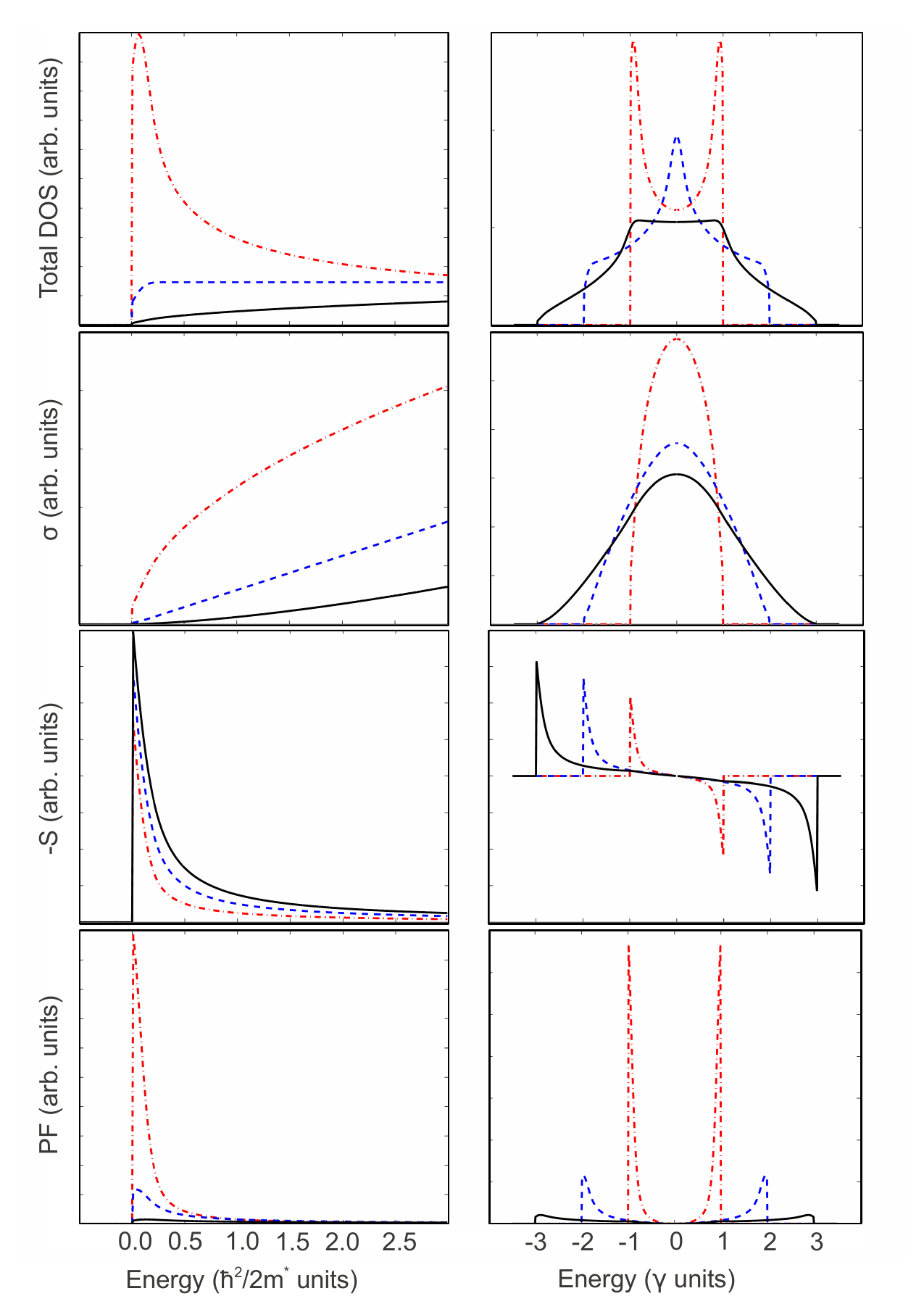}
       \caption{ (Color online) 
                 Density of states (DOS), electrical conductivity ($\sigma$),
                 Seebeck coefficient (changed sign, $-S$) and power factor (PF)
                 for a free electron model as described in Eq.~(\ref{eq:fe})
                 (left panels), and 
                 the tight-binding model of Eq.~(\ref{eq:tb}) (right panels).
                 Different colors represent the dimensionality of the system:
                 solid black lines for three-dimensional,
                 blue dashed lines for two-dimensional,
                 and red dash-dotted lines for one-dimensional.}
        \label{fig:trans-tb}
   \end{center}
\end{figure}

 In Fig. \ref{fig:trans-tb} we compare the DOS and transport properties 
 for both models using the formalism developed in 
 Sec.~\ref{sec:comp}.  For ease of comparison with the  
 (SrTiO$_{3}$)$_5$/(SrRuO$_{3}$)$_1$ superlattice, this plot is structured 
 in a similar way to Fig. \ref{fig:trans}. 

%  DOS

\begin{figure} [h]
    \begin{center}
       \includegraphics[width=7.0cm]{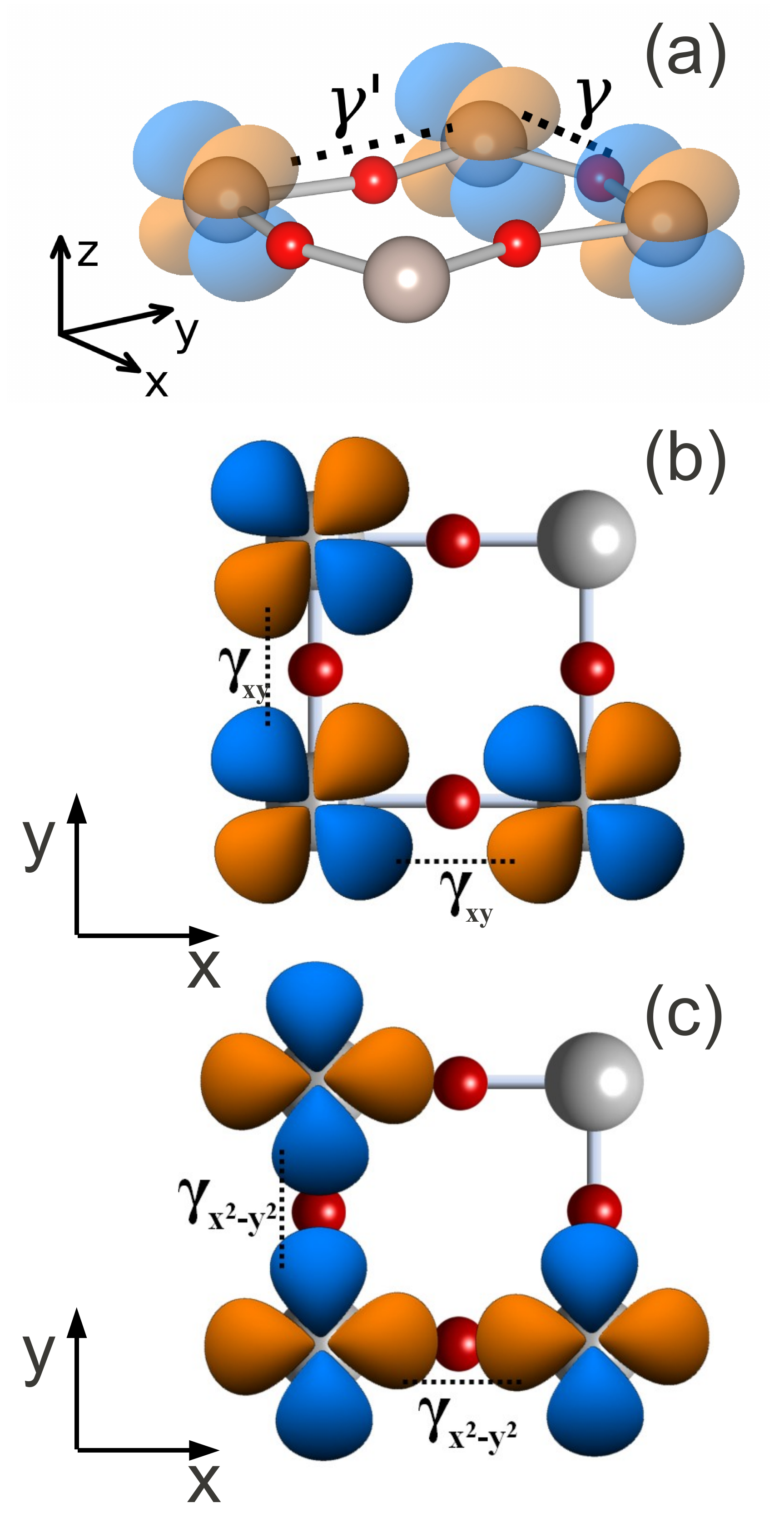}
       \caption{ (Color online) Illustration of the 4$d$ orbitals of Ru 
                 in the RuO$_2$ layer 
                 and hopping parameters, $\gamma$, for (a) the $d_{xz}$ 
                 (an equivalent schema, rotated 90$^\circ$,
                 could be made for the $d_{yz}$),
                 (b) the $d_{xy}$,
                 and (c) the $d_{x^2-y^2}$ orbitals. 
                 In (a), the hopping along $z$ is strongly hindered 
                 as it involves moving into Ti
                 levels which are much higher in energy. 
                 From the overlap of the orbitals it is clear than
                 $\gamma>>\gamma'$.
                 Meaning of the colours for the atoms as in 
                 Fig.~\ref{fig:super-dos}(a).}
        \label{fig:orb}
    \end{center}
 \end{figure}

 The DOS plots correspond with the textbook examples~\cite{Ashcroft,Martin}
 of the corresponding models where, for example, the 2D free-electron model 
 involves a step function. 
 Comparing both models for the same dimensionality we observe that, 
 at the band edge, both are quite similar with 
 a logarithmic divergence (in 1D), 
 a finite discontinuity (in 2D), 
 and a functional dependence with the energy 
 $\propto \sqrt{\varepsilon}$ (in 3D).
 However, the models quickly differ at higher energies. 
 In particular, the tight-binding bands are symmetrical around the  
 center of the band, while the free-electron ones are not. The reason 
 behind the 
 similitudes between the two models comes from the fact that the tight-binding
 dispersion curves can be approximated by parabolae 
 (see Fig. \ref{fig:bands}) at the
 band-edges.
 Comparing these simple models to the first-principles calculated DOS of 
 the conduction band (top-right panel of Fig. \ref{fig:trans}) 
 we find that the most similar one is the 1D tight-binding model.
 In particular, both pictures display a very characteristic two-peaked 
 structure. 
 While it is reasonable that a tight-binding model is more adequate than 
 a free-electron one to describe the narrow 4$d$-bands of Ru, 
 it seems surpring that the DOS resembles that of a 1DEG rather 
 than a 2DEG one. 
 The reason for this behavior is that in the case of 
 Ru(4$d_{xz}$) and Ru(4$d_{yz}$) bands the hopping parameter in the 
 conducting plane is only large along $x$ or $y$ directions, respectively 
 [see Fig.~\ref{fig:orb}(a)]. Thus, the 2DEG in the 
 (SrTiO$_{3}$)$_5$/(SrRuO$_{3}$)$_1$ superlattice is, in fact, formed 
 by two half-filled orthogonal 1D bands. 
 In contrast, in the Ru(4$d_{xy}$) and Ru(4$d_{x^2-y^2}$) bands the 
 orbitals bond equally well in the $x$ and $y$ directions 
 [see Fig.~\ref{fig:orb}(b) and \ref{fig:orb}(c)] 
 forming a proper 2DEG. 
 Indeed, if we compare the DOS of the majority spin Ru(4$d_{x^2-y^2}$) 
 band in Fig.~\ref{fig:tb} we observe a very similar shape to that of 
 the ideal bidimensional tight-binding model shown in 
 Fig.~\ref{fig:trans-tb}. 
 For the Ru(4$d_{xy}$) a good agreement with the model can be also be 
 achieved if the tight-binding expansion is extended to include interactions 
 with in-plane neighbors along $\langle 110 \rangle$ directions that shift  
 the central DOS peak to higher energies. 

% The conductivity

 Regarding the electric conductivity, the results obtained for both 
 the free electron and the tight-binding models are 
 equivalent at the band-edge where the 
 tight-binding bands can be approximated by parabolae. 
 This can be seen in the way the conductivity curves decay in a 
 quicker way as the energy gets closer to the lower bound of the band. 
 However, the behavior for half filling is quite different in both models. 
 In particular, the tight-binding model predicts a maximum conductivity in the 
 middle of the band, a behavior not observed for the free electron 
 approximation. 
 This conductivity maximum corresponds to the maximum of the 
 group velocity at $\varepsilon=0$ as deduced from Eq.~(\ref{eq:tb}) and
 can be graphically visualized as the slope of the band diagram in 
 Fig.~\ref{fig:bands}.
 This maximum can clearly be seen around $\mu=0$ in the full conductivity 
 calculation for the minority spin channel in 
 (SrTiO$_{3}$)$_5$/(SrRuO$_{3}$)$_1$ (Fig.~\ref{fig:trans}), 
 and the 1D character of the band can also be observed in the 
 abrupt reduction of conductivity around $\mu\approx -0.5$ eV.

% Seebeck coefficient

 The Seebeck coefficient follows a similar trend, with the free electron 
 and tight-binding results closely matching each other at the edge 
 of the band.
 At this point, the absolute value of the 
 Seebeck coefficient is larger for bulk (3D) than in lower dimensionality 
 systems (2D, 1D). 
 This may come as a surprise when it was experimentally shown that in the 
 2DEG in Nb-doped SrTiO$_{3}$\cite{Ohta-07.2} the Seebeck 
 constant \emph{increases} with reduced dimensionality. 
 However in that case it was argued~\cite{Ohta-07.2,Choi-10} that 
 polarons are responsible for transport~\cite{Alexandrov-10} 
 and thus the present model cannot account for their behavior. 
 Going back to the comparison between both the free electron and the  
 tight binding models we find that in the former case electronic bands 
 are either for electrons or holes depending on the sign of the curvature in 
 Eq.~(\ref{eq:fe}) around $k$ = 0, while in the latter the bands transit 
 from electron ($S<0$) to hole ($S>0$) dominated with filling. 
 Thus, the value $S$ = 0 is achieved exactly in the middle of the band, 
 as observed both for the model in Fig.~\ref{fig:trans-tb} and the 
 full calculation for the minority spin bands in the full 
 heterostructure (Fig.~\ref{fig:trans}).

% The PF

 Finally, we analyze the power factor in both models. 
 We observe that at the band edge the PF reaches a maximum value that 
 is significantly enhanced when the dimensionality is reduced, 
 in agreement with parabolic band models within effective mass theory
 used by Hicks and Dresselhaus.~\cite{Hicks-93,Hicks-93.2} 
 Taking into account that PF=$S^2\sigma$ it is surprising to find this 
 enhancement knowing that the maximum value of the Seebeck coefficient 
 is reduced with dimensionality. 
 However, this decrease is overcompensated by the increase of the conductivity 
 at the edge of the band. 
 As the decay of the one dimension conductivity is sharper close to 
 the beginning of the band for lower dimensionality systems, 
 the value of the conductivity is also larger close to the edge than in 
 3D leading to a global enhancement of the PF. 
 In the middle of the band the tight-binding model 
 predicts $S$ = 0 which leads to a null value of the PF. 
 Thus, the fact that the conducting band is exactly half filled in 
 (SrTiO$_{3}$)$_5$/(SrRuO$_{3}$)$_1$ superlattices explains their very poor 
 thermoelectric properties even though they display a half-metallic 2DEG with 
 a width of a single SrRuO$_{3}$ layer and a very large thermopower.

%% Conclusions

 Therefore, the strong confinement of the 
 electron gas in (SrTiO$_{3}$)$_5$/(SrRuO$_{3}$)$_1$
 superlattices does not involve a corresponding enhancement of the 
 thermoelectric properties. 
 Recent experiments in the conducting LaAlO$_{3}$/SrTiO$_{3}$ interface
 also point to the absence of an enhancement of the Seebeck coefficient
 due to electronic confinement.~\cite{Pallecchi-10}
 This seems to be at odds with the Hicks and Dresselhauss's
 model.~\cite{Hicks-93,Hicks-93.2} 
 However, there is no such discrepancy: simply, the physical approaches on 
 which the previous mode was based are not fullfilled in the present case.
 The main difference of our system with those of Hicks and Dresselhauss 
 is the fact that the bands in our system are narrow and half-filled,
 mostly due to the potential wells arising predominantly from the ionic 
 charges,~\cite{Mannhart-10}
 breaking the parabolic band approximation used in their calculations.
 
\section{Conclusions}

 In this work we have theoretically studied the transport properties 
 of the 2DEG present in the half-metallic (SrTiO$_{3}$)$_5$/(SrRuO$_{3}$)$_1$ 
 superlattice. Using Boltzmann's transport theory we have shown that: 
 (i) the semiconducting spin channel displays a large 
 Seebeck coefficient, $S \approx$ 1500 $\mu$V/K, larger 
 than that found in Nb-doped 
 SrTiO$_{3}$ thin films,
 (ii) however the total power factor is too small 
 for the system to be a good thermoelectric material. 
 The reason behind the low power factor is the half filling of the 
 metallic spin channel that, while providing a high conductivity, 
 finally quenches the total Seebeck constant of the system.

 Our results do not contradict those of Hicks {\it et al.} 
 \cite{Hicks-93,Hicks-93.2} who predicted an important enhancement 
 of the thermoelectric properties for systems whose transport is 
 strongly confined in one or two dimensions although it shows its limits 
 of application. 
 Hicks and Dresselhaus assumed a semiconducting superlattice with 
 parabolic bands that could be doped to increase the carrier density. 
 However, in our system the main transport bands are narrow as corresponds 
 to those with a strong Ru(4$d_{xz,yz}$) character and are half filled being, 
 as a consequence, not suitable for the application of the free electron model. 
 Indeed, they are well described within the 
 tight-binding approximation. 
 Using this model we show that the 2DEG in (SrTiO$_{3}$)$_5$/(SrRuO$_{3}$)$_1$ 
 is unusual in the sense that is composed of two orthogonal bands 
 where the hopping parameter is only strong in one dimension having thus, 
 the properties of a 1DEG. 

% Acknowledgments

 Financial support from the Spanish Ministery of Science and
 Innovation through the MICINN Grant FIS2009-12721-C04-02, 
 and by the European Union through the project EC-FP7,
 Grant No. CP-FP 228989-2 ``OxIDes''.
 PhG acknowledges financial support from an ARC project TheMoTher,
 IAP project P6/42 from the Belgian state-Belgian Science Policy, and
 through a Research Professorship from the Francqui Foundation.
 D.I.B. acknowledges financial support from the grant of the Romanian
 National Authority for Scientific Research, CNCS-UEFISCDI,
 Project No. PN-II-RU-TE-2011-3-0085.
 The authors thankfully acknowledge the computer resources,
 technical expertise and assistance provided by the
 Red Espa\~nola de Supercomputaci\'on.
 Other calculations were performed on the computers at the ATC group
 of the University of Cantabria and on the NIC3 at ULg.

\end{document}